\documentclass{article} 
\usepackage{iclr2024_conference,times}

\usepackage{hyperref,graphicx,caption,algorithm,algcompatible,multirow,amsmath,booktabs,pifont,diagbox,color}
\usepackage{url}

\newcommand{\cmark}{\ding{51}}%
\newcommand{\cmarkbold}{\ding{52}}
\newcommand{\xmark}{\ding{55}}%

\DeclareMathOperator*{\argmin}{arg\,min}

\title{Everyone Can Attack: Repurpose Lossy Compression as a Natural Backdoor Attack}

\author{%
Sze Jue Yang$^{1,2}$ \quad Quang H. Nguyen$^1$ \quad Chee Seng Chan$^2$ \quad Khoa D. Doan$^1$  \\
$^1$College of Engineering and Computer Science, VinUniversity, Vietnam \\
$^2$Center of Image and Signal Processing, Universiti Malaya, Malaysia\\
jason.y@vinuni.edu.vn, quang.nh@vinuni.edu.vn, cs.chan@um.edu.my, khoa.dd@vinuni.edu.vn \\
}

\iclrfinalcopy 
\begin{document}

\maketitle

\begin{abstract}
The vulnerabilities to backdoor attacks have recently threatened the trustworthiness of machine learning models in practical applications. 
Conventional wisdom suggests that not everyone can be an attacker since the process of designing the trigger generation algorithm often involves significant effort and extensive experimentation to ensure the attack's stealthiness and effectiveness. 
Alternatively, this paper shows that there exists a more severe backdoor threat: anyone can exploit an easily-accessible algorithm for silent backdoor attacks. 
Specifically, this attacker can employ the widely-used lossy image compression from a plethora of compression tools to effortlessly inject a trigger pattern into an image without leaving any noticeable trace; i.e., the generated triggers are natural artifacts. 
One does not require extensive knowledge to click on the ``convert'' or ``save as'' button while using tools for lossy image compression.
Via this attack, the adversary does not need to design a trigger generator as seen in prior works and only requires poisoning the data. 
Empirically, the proposed attack consistently achieves 100\% attack success rate in several benchmark datasets such as MNIST, CIFAR-10, GTSRB and CelebA. More significantly, the proposed attack can still achieve almost 100\% attack success rate with very small (approximately 10\%) poisoning rates in the clean label setting. The generated trigger of the proposed attack using one lossy compression algorithm is also transferable across other related compression algorithms, exacerbating the severity of this backdoor threat. This work takes another crucial step toward understanding the extensive risks of backdoor attacks in practice, urging practitioners to investigate similar attacks and relevant backdoor mitigation methods. 

\end{abstract}

\section{Introduction}
Machine learning, especially deep neural networks (DNNs), has gained popularity due to their superior performance in various applications and tasks such as computer vision \cite{alexnet, resnet}, natural language processing \cite{bert} and healthcare \cite{DBLP:journals/corr/abs-2101-00008}. The emergence of DNNs in high-stake applications has raised security concerns about their vulnerabilities to malicious attacks. Prior research has shown that DNNs are vulnerable to a wide range of attacks, including adversarial examples \cite{cw_loss, madry_adv_examples}, poisoning attacks \cite{poison_backgrad_opt, poison_frog} and backdoor attacks \cite{fed_bd, badnet}. Backdoor attacks impose serious security threats by injecting a stealthy and malicious trigger onto a DNN by poisoning the data or manipulating the training process \cite{neural_trojan, trojan_attack_nn}. The backdoored model will behave normally with clean inputs but behave maliciously whenever the trigger is present in the input. For example, an autonomous vehicle system will normally stop when it encounters a ``stop'' sign, but when a trigger (i.e., a yellow sticker) is present on the sign, the system will misclassify it as ``speed limit of 110'' (the attack target), causing the vehicle to speed up instead of stopping. This scenario demonstrates the severity of backdoor attacks in autonomous vehicle systems, for this malicious behavior may lead to serious accidents.

\begin{figure}[t]
    \centering
    \includegraphics[keepaspectratio, scale=0.35]{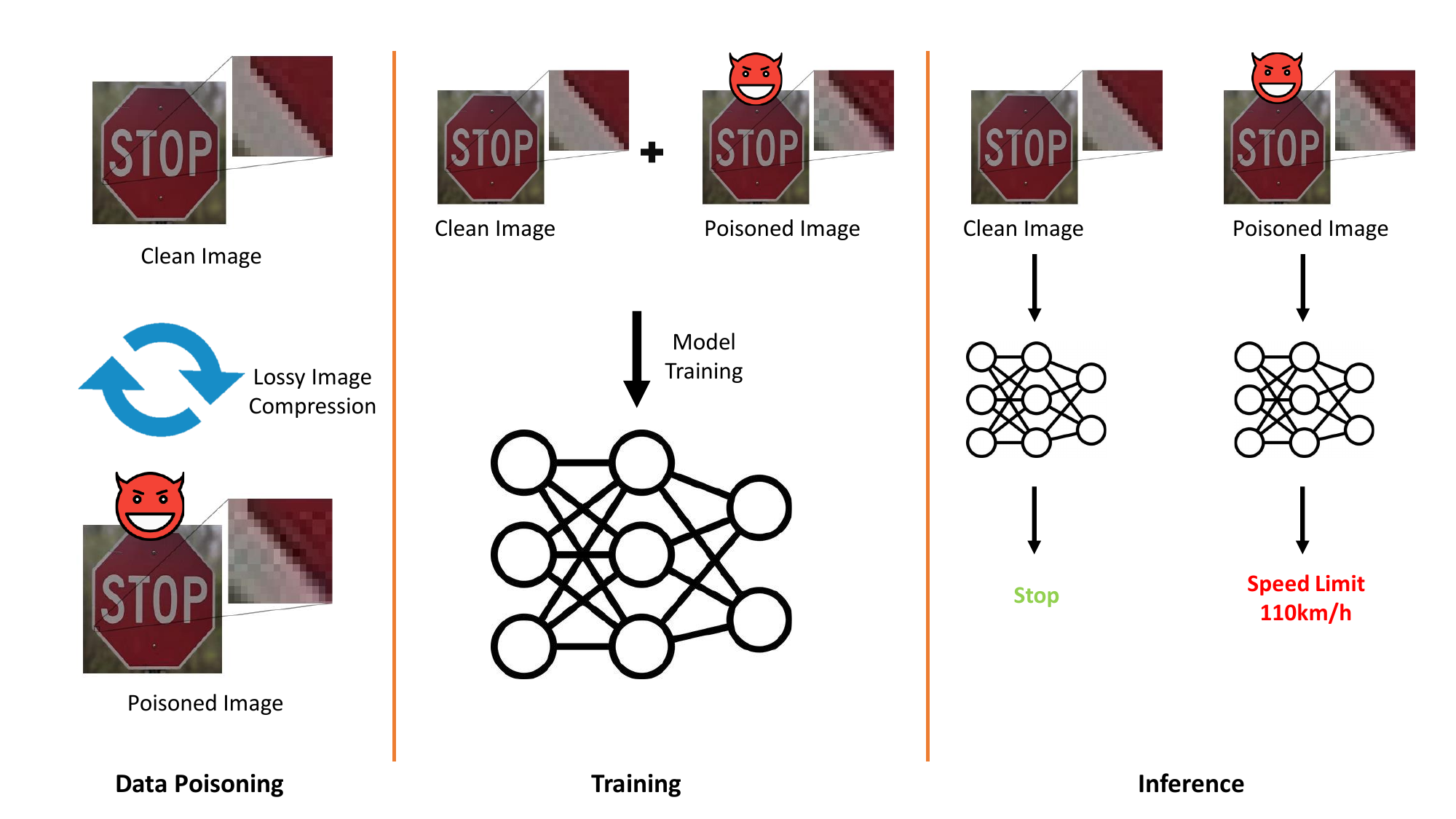}
    \caption{An overview of our attack. In the \textbf{data poisoning} stage, we apply lossy image compression to the original images, and create poison images. During \textbf{training} stage, the clean images and poisoned images are the inputs to the model. For \textbf{inference}, the model behaves normally when we input a clean image, but the backdoor is triggered when the input is a lossy compressed image.}
    \label{fig:figure_1}
\end{figure}

Early works expose the security vulnerability of backdoor attacks via poisoning the training data with hand-crafted but visible triggers~\cite{badnet,liu2020reflection,barni2019new}; when a user trains a model with the poisoned datasets, the backdoor is successfully implanted and the model is under the control of the adversary. Since then, backdoor research has shown various threat models where the adversary can secretly launch an attack. For example, the adversary can design imperceptible triggers to fool both human and machine inspections~\cite{wanet, saha2020hidden}, can arbitrarily control the attack target class during inference~\cite{marksman_bd,shokri2020bypassing}, can hide the backdoor inside model architectures instead of training data~\cite{model_architecture_backdoor,clifford2022impnet}, or can modify the pre-trained weights to craft the attack~\cite{rakin2020tbt,backdoor_cnn}. While the existing works show undeniably harmful types of backdoor attacks, these attacks also demand significant effort and extensive experimentation from the adversary to ensure the attack's stealthiness and effectiveness. For example, WaNet~\cite{wanet} requires a special ``noise''-mode training to ensure its effectiveness, while MAB~\cite{model_architecture_backdoor} requires a special design of a trigger-detecting layer for its attack. Even the classical patch-based~\cite{badnet} or blending-based~\cite{liu2020reflection} triggers require sufficient knowledge of image editing and access to image-editing tools.   
 
This paper continues the quest of unveiling zero-day vulnerabilities in backdoor-attack research by showing that everyone,
not only the competent adversary, can launch backdoor attacks. Specifically, this adversary can re-purpose the widely used algorithm, lossy image compression, and its natural artifacts, a byproduct of the objective to retain the visual similarity between the compressed and original images, to mount an imperceptible and highly-effective backdoor attack for both the dirty-label and clean-label settings. Lossy image compression algorithms are ubiquitous and easily-accessible tools to transfer multimedia resources on the internet or display graphical content in browsers or smart-device applications. Consequently, crafting such an attack requires little or no effort or knowledge of machine learning. The overview of our method are denoted as in Figure~\ref{fig:figure_1}. Our \textbf{contributions} are summarized below:
\begin{itemize}
    \item We make the case that everyone can mount a powerful and stealthy backdoor attack. This sounds the alarm of yet another previously-undiscovered security threat of backdoor attack in practice.
    \item We repurpose a widely-used algorithm, i.e., lossy image compression, to launch this easily-accessible backdoor attack process. The proposed attack process requires little effort to craft the imperceptible trigger and a low poisoning rate to implant the backdoor even in the clean-label setting (which often necessitates a conspicuously higher data-poisoning rate). 
    \item We empirically demonstrate the effectiveness of the proposed attack process on several benchmark datasets and backdoor settings. We show that the proposed method can achieve high attack success rates in both dirty-label and clean-label settings, can bypass most existing defensive measures, and can transfer across related compression algorithms.   
\end{itemize}

\section{Related Works}
\subsection{Backdoor Attacks}
Previous works have formulated backdoor attacks as a process of introducing malicious behaviors to a model, $f_\theta$, parameterized by $\theta$, trained on a dataset, $\mathcal{D}$. 
This process involves a transformation function, denoted as $T(\cdot)$ that injects malicious trigger onto the input, $x$ and form an association with the desired model output, namely the target class, $y_t$ \cite{trojan_attack_nn, badnet, blind_bd}. Currently, the main methodologies to inject this malicious behavior into the model are contaminating the training data \cite{blended, trojan_attack_nn}, altering the training algorithm \cite{blind_bd}, or overwriting/retraining the model parameters \cite{weight_poison_att, backdoor_cnn}.

Discovering zero-day vulnerabilities in DNNs has always been a focus. For instance, BadNet~\cite{badnet} exposed image recognition systems are susceptible to backdoor attacks, by injecting a malicious patch onto the image and changing its label to a predefined target class. Then, HTBA~\cite{saha2020hidden} showed that without changing the labels of a dataset, an adversary can still attack by forming a strong association between a trigger patch and the ground truth class. SIG~\cite{sig} proposed to superimpose sinusoidal waves onto images, whereas ReFool~\cite{liu2020reflection} proposed to create a trigger pattern by the reflection of an image which is mostly imperceptible to human's vision. Besides, WaNet~\cite{wanet} and LIRA~\cite{LIRA} showed that a trigger pattern could be almost invisible from human's perception through optimizing the trigger generation function, shedding lights in the way of another attack. To further elaborate the severity of backdoor attacks, MAB~\cite{model_architecture_backdoor} exploits the model's architecture by adding a pooling layer that will be activated to a trigger pattern, showing a new paradigm of attacking DNNs.

While these methods are effective, extensive experiments and manual interventions are required to design the triggers and verify their effectiveness. In addition, due to the complex nature of the trigger generation process, they are often only available to adversaries with a certain degree of understanding on backdoor attacks. These constraints limit the practicality and severity of the backdoor attacks, as they are only applicable to a small pool of adversaries. 
In contrast, we expose a zero-day vulnerability in backdoor attacks where everyone, not only knowledgeable adversaries, could cause severe damage to DNNs with little to minimal effort needed to design the trigger generator.

\subsection{Backdoor Defense}
Due to the emergence of backdoor attacks, another line of research focusing on preventing and mitigating backdoor attacks has also gained attention. Several works have been developed to counter backdoor attacks such as backdoor detection \cite{activation_clustering, spectral_signature, strip}, input mitigation \cite{neural_trojan, li2020rethinking} and model mitigation \cite{fine_pruning, neural_cleanse}.

Detection-based backdoor defense methods aim to detect backdoored samples by analyzing the model's behavior. For example, 
Activation Clustering \cite{activation_clustering} detects the model's malicious behavior by analyzing the activation values in the latent space and 
STRIP \cite{strip} analyzes the entropy of the model's output on perturbed inputs. Input mitigation methods attempt to remove the trigger of inputs by altering or filtering the image such that the model will retain its normal behavior even when it is backdoored 
(i.e. it suppresses and deactivates the backdoor).

In contrast, model mitigation methods mitigate the backdoor attacks before deployment. Fine pruning \cite{fine_pruning} combines both fine-tuning and pruning to eliminate redundant weights or neurons in DNNs with a training set, hoping to mitigate the injected backdoor.
Besides, Neural Cleanse \cite{neural_cleanse} detects whether a trained model has been backdoored by searching for potential patterns that could trigger the backdoor.

\begin{table}[t]
\centering
\resizebox{0.8\linewidth}{!}{ 
\begin{tabular}{l||c|c|c|c|c} 
\hline\hline
\multicolumn{1}{c||}{Method} & \begin{tabular}[c]{@{}c@{}}Dataset \\Access\end{tabular} & \begin{tabular}[c]{@{}c@{}}Black-box \\Model Access\end{tabular} & Accessible & Natural & Stealthy  \\ 
\hline \hline
BadNet                      & \cmark                                                        & \cmark                                                                & \xmark    & \xmark    & \xmark             \\
WaNet                       & \cmark                                                        & \xmark                                                                & \xmark    & \xmark    & \cmark             \\
LIRA                        & \cmark                                                        & \xmark                                                                & \xmark    & \xmark    & \cmark             \\
 \hline \hline
\textbf{Ours}                        & \cmarkbold                                                        & \cmarkbold                                                         & \cmarkbold       & \cmarkbold       & \cmarkbold             \\
\hline\hline
\end{tabular}}
\caption{Comparison to other methods. ``\cmark'' indicates the attribute is present, while ``\xmark'' indicates the attribute is missing.}
\label{table1:compare}
\end{table}

\section{Threat Model}

Ultimately, a threat model that has all the following five characteristics is the worst nightmare of a backdoor attack. 

\textbf{Dataset Access}: It is defined as the ability to apply modifications onto the model's training dataset. We consider dataset access hazardous, since deep learning projects usually involve data labeling and annotations, creating opportunities for adversaries to act maliciously on the data. 

\textbf{Black-box Model Access}:
It assumes that adversaries are only required to access/poison the data without involvement in the model's training process. 
In practice, this assumption holds as it is usually impossible for an adversary to be involved in the model's training process, as the training recipe and the model's architecture are usually trade secrets. Contrary, since most DNNs are pre-trained using publicly available datasets, it is much easier to poison the public dataset than to access the model.

\textbf{Accessible}: It refers to everyone could become an adversary, even without prior knowledge of machine learning. Specifically, this is the deadly sin of backdoor attack as it refers to anyone who can re-purpose any generic tools to launch a silent backdoor attack. Based on our knowledge, most if not all of the prior work have low accessibility where professional knowledge is a must to drive the complicated trigger generation process to launch an effective attack.

\textbf{Natural}: 
This refers to initiating an attack with a mechanism that is not meant for malicious purposes. The goal of ``natural'' is to reduce the suspicion of backdoor attacks, where the attack objective is shy away from the original intent of the mechanism.

\textbf{Stealthy}: 
It refers to the ability to hide the trigger pattern and backdoor attack from human inspections (i.e. poisoned images have high visual similarity to the clean image). Also, it is desirable for an attack to be stealthy against machine inspection, i.e., defensive algorithms. However, bypassing all existing defensive algorithms is extremely challenging and even impossible in the backdoor domain~\cite{DBLP:journals/corr/abs-2007-08745}

On the whole, fortunately or unfortunately, Table \ref{table1:compare} shows this ``exorcist'' threat model has not been discovered until now. In next section, we show that the widely-used lossy image compression can be re-purposed as a natural backdoor attack. Also, since image compression is common and countless compression tools are available online/offline, everyone can easily launch a backdoor attack. This discovery takes another crucial step toward understanding the extensive risks of backdoor attacks in practice.

\section{Methodology}
\subsection{Problem Formulation}
Consider a supervised image classification task, where a model, $f_\theta$ maps the input domain, $\mathcal{X}$ onto the target classes, $\mathcal{C}$, where $\theta$ is the trainable parameters: $f_\theta : \mathcal{X} \rightarrow \mathcal{C}$. The main objective is to learn $\theta$ from the training dataset $\mathcal{D} = \{(x_i, y_i) : x_i \in \mathcal{X}, y_i \in \mathcal{C}, i = 1, \cdots, N \}$. 

In backdoor attacks, a model, $f_\theta$ is trained with the combination of both clean and poisoned subsets of $\mathcal{D}$. Technically, in order to create a poisoned subset from the dataset, $\mathcal{D}$, a clean sample with corresponding label, $(x, y)$ is transformed into a backdoor sample $(T(x), \eta(y))$, where $T$ is a backdoor transformation function that converts a benign input, $x$ into a poisoned input, $\hat{x}$. $\eta$ is a target transform function, which converts an original class to other target classes. When training $f_\theta$ with clean and poisoned samples, the behavior of $f_\theta$ is altered such that: \[f_\theta(x) = y, \quad f_\theta(T(x)) = \eta(y), \tag{1}\] for every pair of clean data $x \in \mathcal{X}$ and its corresponding label, $y \in \mathcal{C}$. 

Generally, there are three commonly studied backdoor attacks: (i) \emph{all-to-one}, (ii) \emph{all-to-all} and (iii) \emph{clean label}. In {\bf all-to-one attack}, the true label is changed to a constant target, $\eta(y) = c, c \in \mathcal{C}$; while in {\bf all-to-all attack}, the true label is one-shifted, $\eta(y) = (y + 1) \, mod \, \vert \mathcal{C} \vert, y \in \mathcal{C}$. In contrast, {\bf clean label attacks} do not change the true label, and only apply $T$ onto the target class' images, with the hope that the trigger pattern will create a strong association to the true label, causing misclassification when the trigger is applied to images from other classes, i.e. $f_\theta(T(x_{c_1})) = y_1, \, f_\theta(T(x_{c_2})) = y_1$ where $c_1 \neq c_2; c_1, c_2 \in \mathcal{C}$. 

\begin{figure}[t]
    \centering
    \includegraphics[keepaspectratio=true, scale = 0.52]{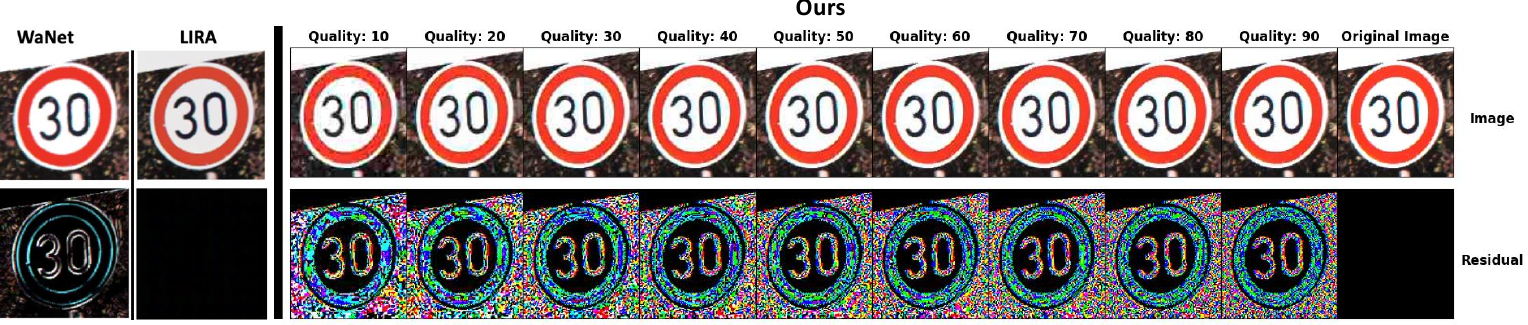}
    \caption{Residual of WaNet, LIRA and our method. Our method has larger magnitude of residuals compared to both WaNet and LIRA. The lower the quality of compressed image, the stronger the magnitude of the residuals, the higher the compression rate. Note that residuals of our method are \emph{natural artifacts} generated lossy image compression algorithms.}
    \label{fig:residual}
\end{figure}

Existing works have been focused on engineering an effective trigger transformation function, to achieve a high attack success rate and stealthiness. However, this process involves extensive design and engineering, and therefore very time-consuming and resource-intensive. Moreover, prior works also assume that the adversaries have access to the data and the model's training process. This assumption poses constraints on the threat model, as it does not reflect real world scenarios. For instance, the adversaries usually have zero to no access to the model's training process. 

\subsection{Trigger Generation}
This paper shows a zero-day vulnerability where anyone can exploit any easily-accessible algorithm for potential silent backdoor attacks. As such, an attacker does not need to engineer a trigger generator as in prior works~\cite{badnet,wanet,LIRA} and only requires poisoning the data. Specifically, we show that lossy image compression can be re-purposed as the transformation function, $T$ as it fulfills the following criteria: (i) \emph{accessible}, (ii) \emph{natural} and (iii) \emph{stealthy}, as shown in Table~\ref{table1:compare}.

\textbf{Accessible:} We show that by re-purposing the lossy image compression algorithm, everyone could now become an adversary as easily as clicking a ``convert'' or ``save as'' button. This is because lossy image compression is widely accessible via a plethora of compression tools on the Internet, such as PNG to JPEG converter, or in a local machine (e.g. MS Paint or Adobe Photoshop). As such, this shows that little or no effort or knowledge of machine learning is required to launch a backdoor attack successfully. To the best of our knowledge, this is the first work that investigates and attains the accessibility of backdoor attacks.

\textbf{Natural:} The idea of lossy image compression algorithms is to compress the image information in the chrominance channel where human visual systems are naturally less sensitive \cite{BULL201417}. As such, we could view the byproduct of lossy image compression as a natural backdoor trigger. That is, to launch a backdoor attack, an adversary can maliciously inject the natural artifacts as a trigger pattern into an image without leaving any noticeable trace.

\textbf{Stealthy}: The original goal of lossy image compression is to reduce image size, while preserving the image contents visually. Naturally, it ensures the trigger's imperceptibility; hence, we could guarantee the stealthiness of our trigger visually. In terms of machine inspection, we show in Sec. \ref{sec:defense}, our proposal is resilient against some popular backdoor defensive algorithms.

Accordingly, our embarrassingly simple, but deadly threat model can be formulated as follows:
\begin{equation}
    T(x) = C_p(x) \tag{2}
\end{equation}
where $C(\cdot)$ could be any publicly available lossy image compression algorithms, $p \in \{1, \cdots, n\}$. We choose JPEG compression \cite{jpeg2} and WEBP \cite{webp} in this paper, as they are among the most widely used.

\subsection{Trigger Injection}
Consider the empirical risk minimization setting where one hopes to minimize the following loss function:
\begin{equation}
    \theta^* = \argmin_\theta \sum_{i=1}^N \mathcal{L}(f_\theta(x_i), y_i) \tag{3}
\end{equation}
Our goal is to minimize the risks, to yield an optimal classifier, $f_\theta$ that could map $x_i$ to $y_i$ correctly. As our transformation function is non-trainable, we apply our transformation function directly to the data, and create a poisoned subset. We formulate the empirical risk minimization objective with backdoor samples as follows:
\begin{equation}
    \theta^* = \argmin_\theta \sum_{i=1}^N \mathcal{L}(f_\theta(x_i), y_i) + \sum_{j=1}^M \mathcal{L}(f_\theta(T(x_j)), \eta(y_j))
    \tag{4}\label{eq5}
\end{equation}
where $N$ is the total number of clean images, $M$ is the total number of poisoned images. By optimizing Eq.~\eqref{eq5}, we are able to jointly optimize the model, $f_\theta$ for both benign and backdoor samples. 

\begin{figure}[t]
    \centering
    \includegraphics[keepaspectratio, scale=0.27]
    {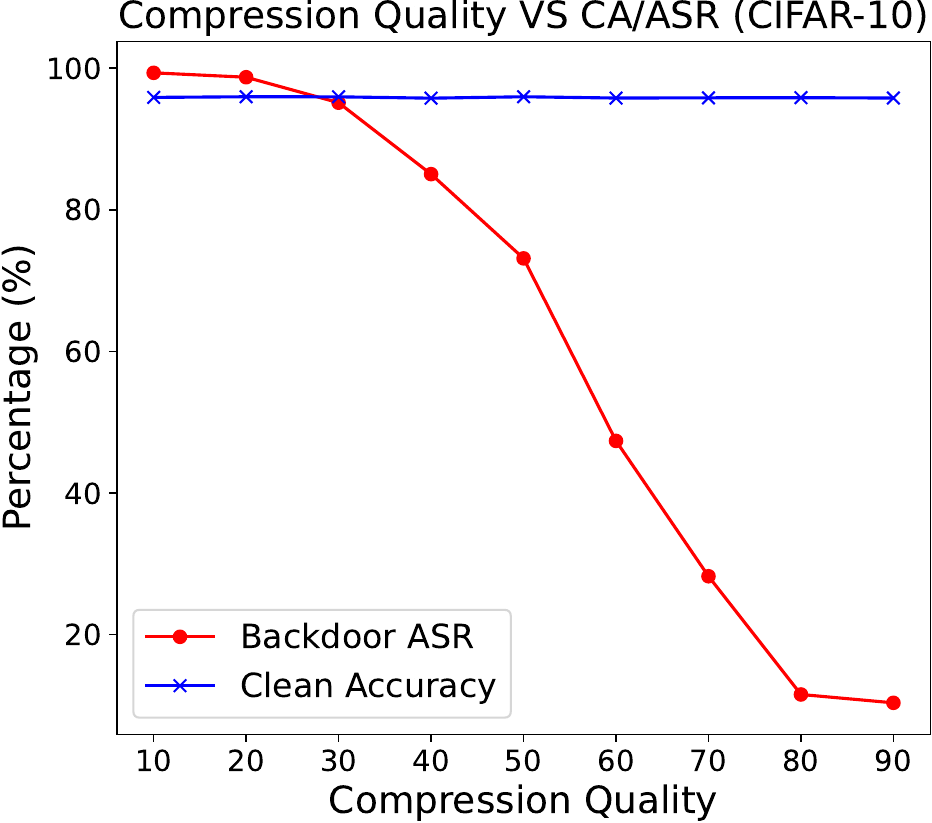}
    \caption{Compressed image quality vs. CA/ASR under 0.1\% poisoning rate. At compression quality of 30, the poisoned image still retain high visual similarity with the original, yet it is able to achieve near 100\% ASR.}
    \label{fig:cifar10_0001_quality_ca_asr}
\end{figure}

\begin{table}[t]
\centering
\begin{tabular}{l||l||c|c|c|c} 
\hline\hline
\multicolumn{1}{c||}{\diagbox{Attack}{Dataset}} & \multicolumn{1}{c||}{} & \begin{tabular}[c]{@{}c@{}}MNIST \\(PNG)\end{tabular} & \begin{tabular}[c]{@{}c@{}}CIFAR-10 \\(PNG)\end{tabular} & \begin{tabular}[c]{@{}c@{}}GTSRB \\(PPM)\end{tabular} & \begin{tabular}[c]{@{}c@{}}CelebA \\(JPEG)\end{tabular} \\ 
\hline \hline
\multirow{2}{*}{WaNet}                         & All-to-One            & \textcolor{blue}{0.99 }/ \textcolor{red}{0.99}        & \textcolor{blue}{0.94~}/~\textcolor{red}{0.99}           & \textcolor{blue}{0.99~}/~\textcolor{red}{0.98}        & \textcolor{blue}{0.79~}/~\textcolor{red}{0.99}          \\ 
\cline{2-6}
                                               & All-to-All            & \textcolor{blue}{0.99~}/~\textcolor{red}{0.95}        & \textcolor{blue}{0.94~}/~\textcolor{red}{0.93}           & \textcolor{blue}{0.99~}/~\textcolor{red}{0.98}        & -                                                       \\ 
\hline
\multirow{2}{*}{LIRA}                          & All-to-One            & \textcolor{blue}{0.99~}/ \textcolor{red}{1.00}        & \textcolor{blue}{0.94~}/ \textcolor{red}{1.00}           & \textcolor{blue}{0.99~}/ \textcolor{red}{1.00}        & -                                                       \\ 
\cline{2-6}
                                               & All-to-All            & \textcolor{blue}{0.99~}/~\textcolor{red}{0.99}        & \textcolor{blue}{0.94~}/~\textcolor{red}{0.94}           & \textcolor{blue}{0.99~}/ \textcolor{red}{1.00}        & -                                                       \\ 
\hline \hline
\multirow{2}{*}{\textbf{Ours @ 5\%}}                  & All-to-One            & \textcolor{blue}{0.98~}/~\textcolor{red}{0.99}        & \textcolor{blue}{0.96~}/ \textcolor{red}{1.00}           & \textcolor{blue}{0.97~}/ \textcolor{red}{1.00}        & \textcolor{blue}{0.80~}/ \textcolor{red}{1.00} \\ 
\cline{2-6}
                                               & All-to-All            & \textcolor{blue}{0.98~}/~\textcolor{red}{0.95}        & \textcolor{blue}{0.96~}/~\textcolor{red}{0.87}           & \textcolor{blue}{0.97~}/~\textcolor{red}{0.91}        & \textcolor{blue}{0.80~}/~\textcolor{red}{0.76}\\ 
\hline
\multirow{2}{*}{\textbf{Ours @ 1\%}}                  & All-to-One            & \textcolor{blue}{0.98~}/~\textcolor{red}{0.88}        & \textcolor{blue}{0.96~}/ \textcolor{red}{1.00}           & \textcolor{blue}{0.97~}/~\textcolor{red}{0.99}        & \textcolor{blue}{0.80~}/ \textcolor{red}{1.00} \\ 
\cline{2-6}
                                               & All-to-All            & \textcolor{blue}{0.99~}/~\textcolor{red}{0.01}        & \textcolor{blue}{0.96~}/~\textcolor{red}{0.81}           & \textcolor{blue}{0.97~}/~\textcolor{red}{0.73}        & \textcolor{blue}{0.80~}/~\textcolor{red}{0.74} \\
\hline\hline
\end{tabular}
\caption{Attack Results. \textcolor{blue}{Blue} denotes the \textcolor{blue}{Clean Accuracy (CA)}, while \textcolor{red}{red} denotes the \textcolor{red}{Attack Success Rate (ASR)}. ``-'' denotes that the result is not available from the original paper. Note that both WaNet and LIRA use 10\% as their poisoning rate; while we achieved comparable results with much lower poisoning rates (i.e. 5\% and 1\%, respectively).}
\label{tab1:att}
\end{table}

\begin{table}[t]
\centering
\begin{tabular}{l||c|c||c|c} 
\hline\hline
\multicolumn{1}{c||}{\multirow{2}{*}{Dataset }} & \multicolumn{2}{c||}{JPEG to WEBP} & \multicolumn{2}{c}{WEBP to JPEG}  \\ 
\cline{2-5}
\multicolumn{1}{c||}{}                          & Clean                 & Attack                  & Clean                 & Attack                  \\ 
\hline \hline
MNIST                                          & 0.98                  & 0.43                    & 0.98                  & 1.00                    \\ 

CIFAR-10                                       & 0.96                  & 0.41                    & 0.96                  & 0.92                    \\ 

GTSRB                                          & 0.97                  & 0.97                    & 0.97                  & 1.00                    \\ 

CelebA                                         & 0.80                  & 0.42                    & 0.80                  & 0.86                    \\ 

\hline\hline
\end{tabular}
\caption{All-to-One Attack Transferability}
\label{tab3:tranferability-all2one}
\end{table}

\section{Experimental Results}

\subsection{Experimental Setup}
We choose four widely used datasets for backdoor attack studies: \textbf{MNIST} \cite{MNIST}, \textbf{CIFAR-10} \cite{CIFAR10}, \textbf{GTSRB} \cite{GTSRB} and \textbf{CelebA} \cite{CelebA}. For the classifier, $f$, we follow WaNet and LIRA, where we used the same CNN for MNIST, Pre-Activation ResNet-18 \cite{resnet} for both CIFAR-10 and GTSRB, and ResNet-18 for CelebA. 

For attack experiments, we compare our results against WaNet and LIRA, as they achieved state-of-the-art results. For hyperparameters, our initial learning rate is 1e-6, and we increase the learning rate to 5e-4 after 5 epochs. We use AdamW as our optimizer and follow a cosine learning rate schedule. We trained our classifier for 300 epochs. For the batch size, we used 1024 for all the datasets. We used a \textbf{lower poisoning rate of 5\%} across our experiments. For the augmentation settings, we follow WaNet and LIRA. We remain the settings across all experiments, and conduct experiments in Pytorch~\cite{pytorch}.

For the lossy image compressions, we used two libraries: Pillow \cite{clark2015pillow} and OpenCV \cite{opencv_library}. 
OpenCV allows specifying compression quality of the image. By default, we use Pillow for our experiments. We found that Pillow's compression is equivalent to OpenCV's compression quality of 75. We conduct experiments on JPEG and WEBP compressions, as they are commonly used. We use JPEG compression by default.

MNIST and CIFAR-10 are naturally stored without compression, therefore we treat them as in .PNG format. For GTSRB, it is stored in .PPM format, which is a lossless image compression algorithm. For CelebA, the images are stored in .JPEG extension. We clarify that even if the images are in lossy compression format, it is possible to re-compress them with lossy image compression to create the triggers.

\begin{table}[t]
\centering
\begin{tabular}{l||c|c||c|c} 
\hline\hline
\multicolumn{1}{c||}{\multirow{2}{*}{Dataset }} & \multicolumn{2}{c||}{JPEG to WEBP}             & \multicolumn{2}{c}{WEBP to JPEG}              \\ 
\cline{2-5}
\multicolumn{1}{c||}{}                          & Clean                 & Attack                & Clean                 & Attack                \\ 
\hline \hline
MNIST                                          & 0.98                  & 0.19                  & 0.99                  & 0.00                  \\ 

CIFAR-10                                       & 0.96                  & 0.26                  & 0.96                  & 0.81                  \\ 

GTSRB                                          & 0.97                  & 0.65                  & 0.97                  & 0.90                  \\ 

CelebA                                         & 0.81                  & 0.02                  & 0.80                  & 0.45                  \\ 

\hline\hline
\end{tabular}
\caption{All-to-All Attack Transferability}
\label{tab4:transferability-all2all}
\end{table}

\subsection{Attack Experiments}
In this experiment, we demonstrate that everyone can launch a backdoor attack, while achieving both attack effectiveness and stealthiness. First, we poison the classifier for each compared backdoor method and calculate its Clean Accuracy (CA) and Attack Success Rate (ASR). CA measures the model's accuracy without any trigger, while ASR measures the model's accuracy on poisoned images. We train and evaluate the backdoor models in three different settings: \emph{all-to-one}, \emph{all-to-all} and \emph{clean label}. We assume that the adversary only has access to the data, but is not involved in the model training phase. As such, our threat model is relatively relaxed compared to both WaNet and LIRA.
Overall, we observe stronger residuals generated by our method compared to both WaNet and LIRA in Figure~\ref{fig:residual}. Even our method creates a larger magnitude of residuals, the trigger remains stealthy. We study and analyze the relationship between stronger trigger strength (i.e. lower compression quality) and CA/ASR.

\subsubsection{All-to-One Attack}
We evaluate our method by setting the target class to a constant class, which is 0. Table~\ref{tab1:att} shows that our method can achieve comparable performance against WaNet and LIRA, even though we used a \emph{much lower poisoning rate of 1\% and 5\%}. 
We further investigate the effectiveness of our method by selecting a significantly lower poisoning rate, 0.1\% ($\sim$50 samples) and evaluate with varying compression quality on CIFAR-10. The results are shown in Figure~\ref{fig:cifar10_0001_quality_ca_asr}. We observe that the ASR is inversely proportional to the compression quality, depicting that a lower compression quality would produce a strong trigger pattern (i.e. large magnitude of artifacts), leading to higher ASR. Given the freedom of varying compression quality, adversaries can select the desired compression quality, causing different magnitudes of damage to DNNs.

Therefore, our method can achieve comparable ASR even with a significantly lower poisoning rate (\textbf{10x lower}).
The trigger generated through lossy image compression algorithms has greater magnitudes of residuals compared to both WaNet and LIRA, as shown in Figure~\ref{fig:residual}, but the trigger remains stealthy to humans.
\begin{figure}[t]
    \centering
    \includegraphics[keepaspectratio=True, scale=0.27]{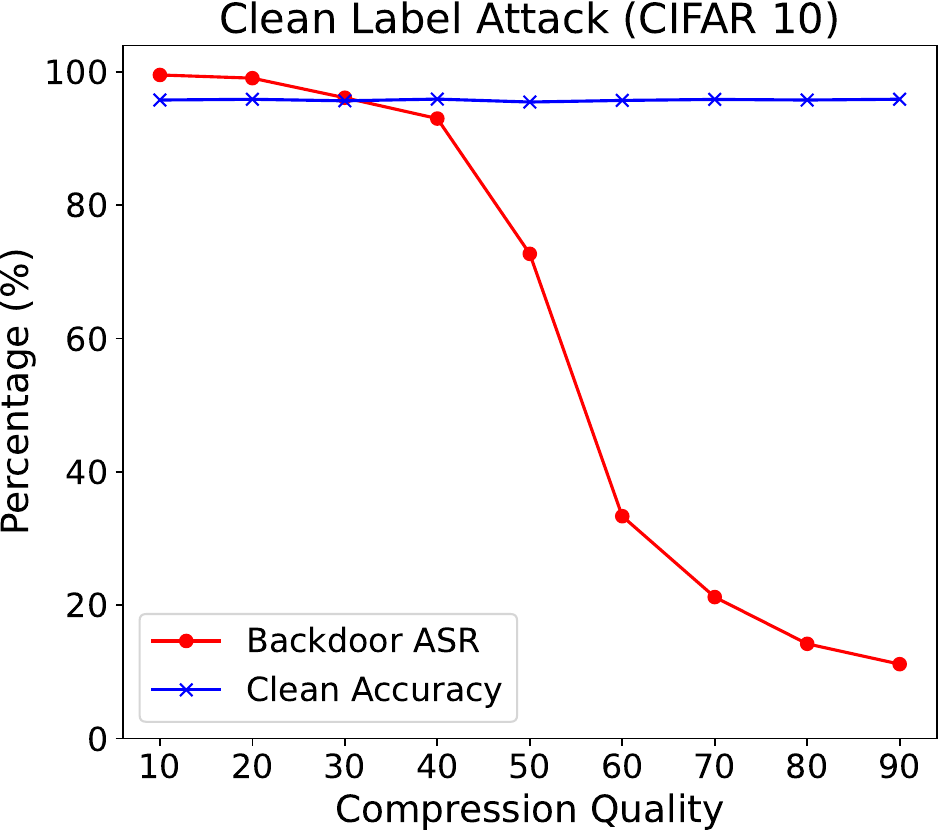}
    \caption{Clean Label Attack on CIFAR-10. We show that compression quality of 30 are able to achieve near 100\% ASR, while remaining stealthy.}
    \label{fig2:cifar10_clean_label}
\end{figure}

\subsubsection{All-to-All Attack}
Under this setting, the true label is converted to the target label by one-shifting. This attack aims to introduce a malicious behavior to a model where a backdoor trigger leads to the prediction of different classes, instead of a fixed target class. It resembles forming a one-to-many relationship between a trigger and multiple classes. Although \emph{our poisoning rate is lower (5\%)}, we observe a comparable result against WaNet and LIRA on all datasets as shown in Table~\ref{tab1:att}.

\subsubsection{Clean Label Attack}
In a clean label attack, the target labels remain the same. Instead, we only poison the images of the target class. By poisoning the target class' images, we create a malicious association between the trigger and the target class. Therefore, when the trigger is applied to other class' samples, the malicious association will 
mislead the model's predictions onto the target class.
We poison \emph{only 10\% of the target class 0 in CIFAR-10}.

\begin{figure}[t]
    \centering
    \includegraphics[keepaspectratio=True, scale=0.27]{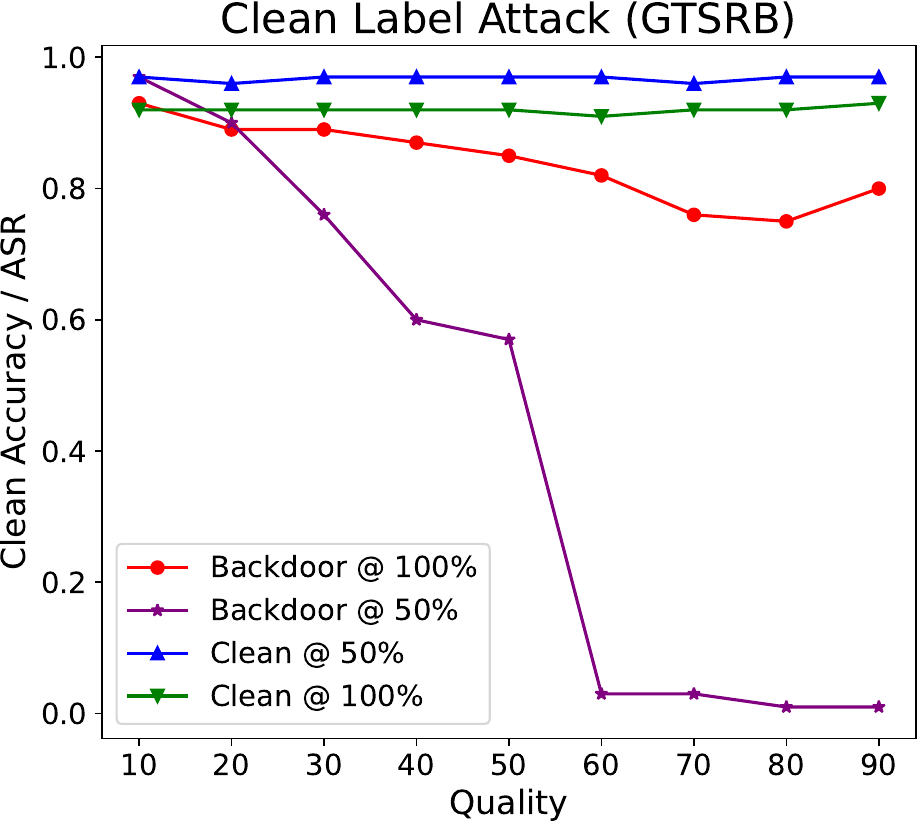}
    \caption{Clean Label attack on GTSRB. We show that given the poisoning rate is high (100\%), even high compression quality could achieve high ASR, while if poisoning rate is lower (50\%), it requires lower compression quality (20) to achieve near 100\% ASR.}
    \label{fig:gtsrb_clean_label}
\end{figure}

In Figure \ref{fig2:cifar10_clean_label}, we showed that our method could achieve nearly 100\% ASR when the compression quality is 30 while remaining stealthy, as shown in Figure~\ref{fig:residual}. Similarly, we evaluate GTSRB (see Figure \ref{fig:gtsrb_clean_label}), with 50\% and 100\% poisoning rates of class 1, respectively. We observe a similar trend where our method achieves near 100\% ASR in GTSRB, while preserving the stealthiness of our trigger. The effectiveness of our attack is due to the
widespread of our trigger across the image; i.e. lossy image compression algorithms will create artifacts on the entire image, disrupting the originally embedded features and linking the samples to the targeted class.
 
On top, our approach also offers an additional option, i.e. set out the image's compression quality. This allows adversaries a certain degree of freedom to compensate for the tradeoffs between ASR and stealthiness. We observe that ASR is inversely proportional to the compression quality, i.e. the trigger strength increases proportionally as the compression quality decreases (see Figures~\ref{fig2:cifar10_clean_label}-\ref{fig:gtsrb_clean_label}).

\begin{figure}[t]
\centering
    \includegraphics[keepaspectratio=true, scale = 0.24]{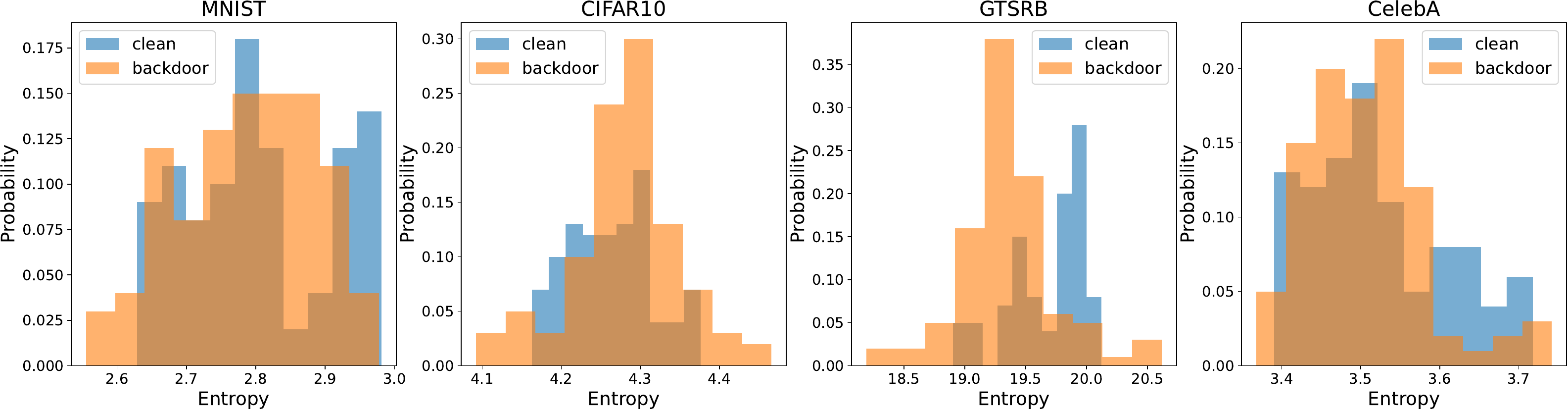}
    \caption{STRIP. We observe similar entropy range between the backdoored model and the clean model, as our trigger combines with the features of the images, which will be disrupt by perturbations.}
    \label{figure3:strip}
\end{figure}
\begin{figure}[t]
\centering    
    \includegraphics[keepaspectratio=true, scale = 0.24]{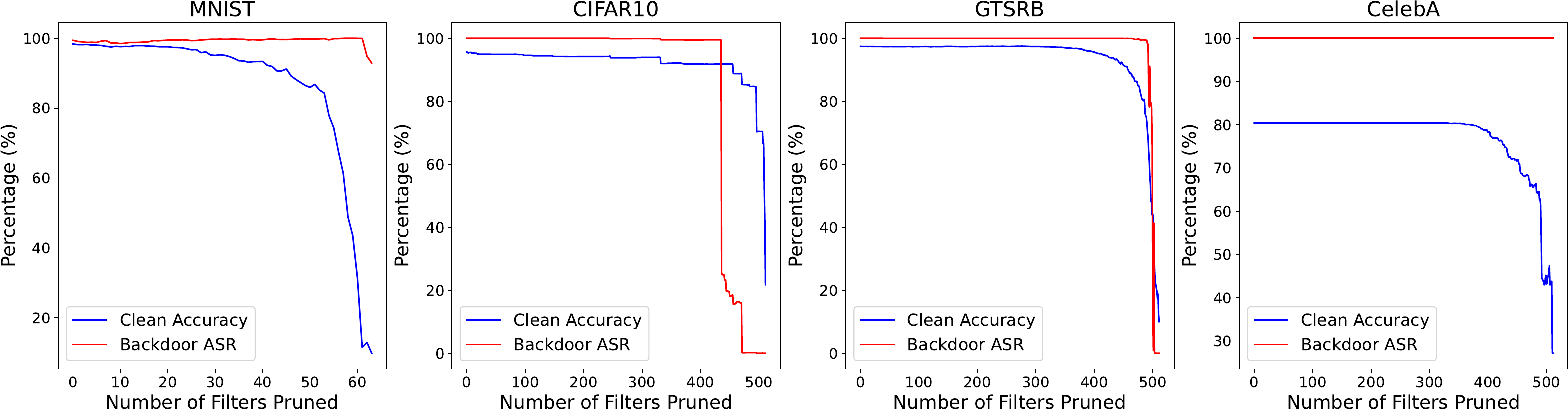}
    \caption{Fine Pruning. We observe that even high number of neurons are pruned, our ASR remains high, depicting the resistance of our backdoor trigger to pruning.}
    \label{figure4:fp}
\end{figure}

\subsubsection{Transferability of Attack}
We evaluate the transferability of our attack across different lossy image compression algorithms (see Table~\ref{tab3:tranferability-all2one}-\ref{tab4:transferability-all2all}). We train a model with JPEG trigger, and then evaluate it with WEBP trigger and vice versa. We show that even if a model is trained with JPEG compression, it is still susceptible to WEBP compression's attack, revealing another possibility of attacking mechanism. 

The results are presented in Tables \ref{tab3:tranferability-all2one}-\ref{tab4:transferability-all2all}, respectively. 
We observe that WEBP trigger has better transferability compared to JPEG. This is because WEBP is a better lossy image compression algorithm \cite{webp} where it tries to preserve more information content of an image (i.e. smaller artifacts), while reducing the image's file size. 

Therefore, WEBP-compressed images have better stealthiness and the artifacts created are harder to learn by the model. Once the model learns the artifacts created by WEBP, it could be attacked even with JPEG-compressed images. We do not conduct experiments on transferability from lossy to lossless image compression algorithms as the nature of lossless image compression algorithms is to preserve all information content. Therefore, lossless image compression algorithms will not generate any artifacts that could be used as a trigger pattern.

\begin{figure}[t]
    \centering
    \includegraphics[keepaspectratio=true, scale = 0.27]{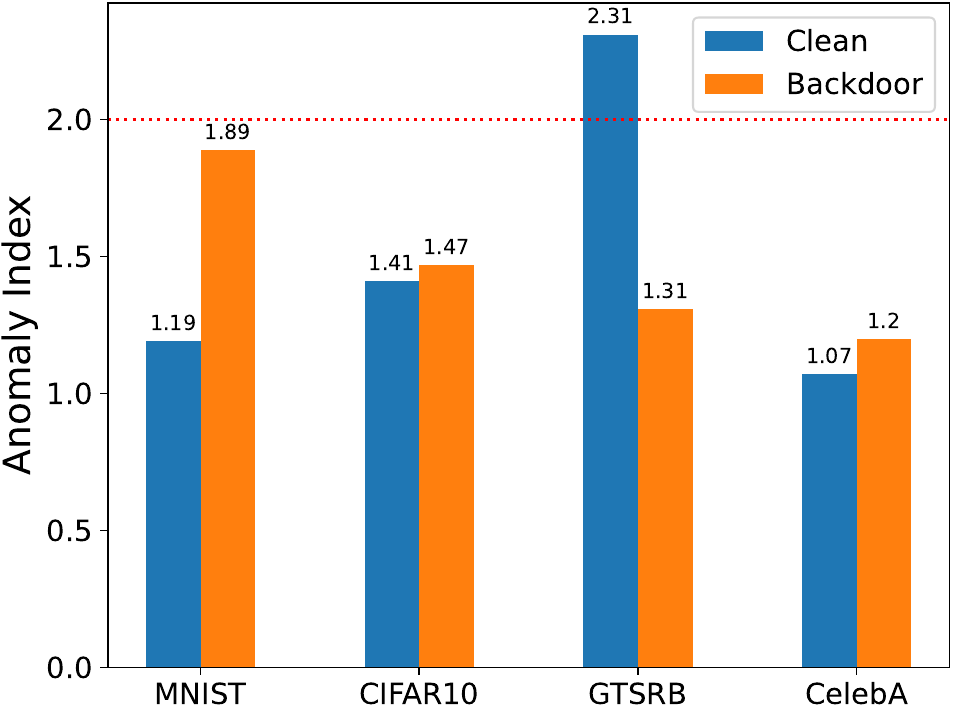}
    \caption{Neural Cleanse. This shows that our method is not detectable as Anomaly Index is $<$ 2 for all backdoor models.}
    \label{figure2:neural_cleanse}
\end{figure}

\begin{figure}[t]
\centering
    \includegraphics[keepaspectratio=True, scale=0.55]{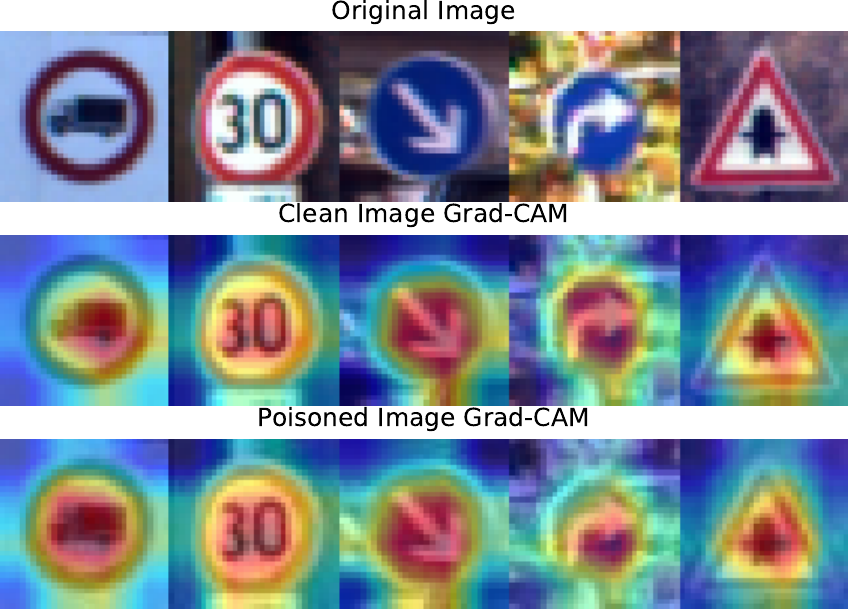}
    \caption{Grad-CAM for GTSRB. This shows that the model has similar activation maps on images with/without trigger, inferring that it ``looks'' at similar region during inference.}
    \label{figure5:gradcam}
\end{figure}

\subsection{Defense Experiments}
\label{sec:defense}
In this section, we evaluate the backdoor-injected classifiers against several popular backdoor defense mechanisms, such as STRIP \cite{strip}, Fine Pruning \cite{fine_pruning}, Neural Cleanse \cite{neural_cleanse} and Grad-CAM \cite{gradcam}.

\subsubsection{STRIP}
We first evaluate our method against STRIP where it perturbs a small subset of clean images and calculates the entropy of the model's prediction. A high entropy in STRIP indicates a low possibility of backdoor attacks, and vice versa. STRIP assumes that a poisoned image will create a strong association between the trigger and the target. Therefore, if the model does not change its prediction with perturbed inputs, the model has a high chance of being backdoored. Figure \ref{figure3:strip} shows that our method has comparable entropy to the clean model across all datasets. With our method, the trigger is implanted across the entire image and ``combined'' with the original image content; hence, perturbations by STRIP will break the trigger along with the original image content.
Therefore, our method behaves like genuine models with similar entropy ranges.

\subsubsection{Fine Pruning}
Fine Pruning analyzes the response of neurons at a specific layer. Given a set of clean images, Fine Pruning feeds the clean images into the network. It identifies the less active neurons, assuming that they are closely related to the trigger, and will not be activated by clean images. To mitigate the backdoor attack, this method works by pruning these dormant neurons, to reduce the effect of backdoor attacks on the network. In Figure~\ref{figure4:fp}, we found that our method is resilient to Fine Pruning in all datasets. This is because our trigger pattern will activate the neurons evenly, causing the backdoor neurons and benign neurons to have indistinguishable responses towards clean images.

\subsubsection{Neural Cleanse}
Neural Cleanse is a model-defense method based on the pattern optimization approach. Neural Cleanse computes the optimal patch pattern that converts the prediction of a clean input to a target class. Then, it checks if any label has a significantly smaller pattern, which is a sign of a backdoor. Neural Cleanse classifies whether a model is backdoored by using Anomaly Index metric, $\tau$ with a threshold $\tau = 2$. A model is considered as backdoored when $\tau > 2$, and vice versa. We ran Neural Cleanse across all datasets and collect the Anomaly Index as shown in Figure~\ref{figure2:neural_cleanse}. We observe that our attack is not detected across all datasets. We also observe that GTSRB has an abnormally high Anomaly Index in a clean model. In CIFAR-10, and CelebA, our method has almost similar Anomaly Index as the clean model.

\subsubsection{Grad-CAM}
To further show the effectiveness of our backdoor attack, we employ visualization from Grad-CAM \cite{gradcam} to understand the backdoored network behavior on both clean and poisoned images. A patch-based trigger could be exposed by Grad-CAM easily as it occupies a small region in the image. In contrast, our method will create a trigger pattern across the image, making it undetectable by Grad-CAM. As shown in Figure \ref{figure5:gradcam}, the heatmaps of our method look similar to the clean model.

\section{Conclusion}
This paper shows everyone could become an adversary against deep learning systems by re-purposing the lossy image compression, which is easily-accessible and widely-available. As such, our method requires minimal to no efforts/knowledge in designing an effective trigger generator. Besides, we proved that the accessibility of backdoor attacks seriously threatens all deep learning systems with extensive experiments. Specifically, we show that not only does the attack remain stealthy, but everyone can now launch such an attack easily. To the best of our knowledge, we are the first to investigate the accessibility of backdoor attacks. We urge researchers and practitioners to investigate this new line of attack, given the potentially serious consequences it could bring to deep learning systems.

\bibliography{aaai24}
\bibliographystyle{iclr2024_conference}

\end{document}